\documentclass[english,phd]{us-phdth}

\usepackage{float}
\usepackage{subcaption}
\usepackage{enumitem}
\usepackage{amsmath,amsfonts,amssymb}
\usepackage{nicematrix}

\usepackage[hypertexnames=false]{hyperref}

\usepackage[%
backend=biber,%
style=phys,%
autocite=plain,%
articletitle=true,biblabel=brackets,%
chaptertitle=true,pageranges=false,%
defernumbers=true
]
{biblatex}
\author{Mateusz Wi\'{s}niewski}
\supervisor{Jakub Spiechowicz, PhD, DSc, Associate Professor}
\title{Effective mass approach to dynamics of non-Markovian systems\\with short memory}

\begin{document}

\frontmatter
\maketitle
{
\pagestyle{empty}
\phantomsection
\chapter*{Abstract}
\addcontentsline{toc}{chapter}{Abstract}

Memory is an inherent feature of the dynamics of physical systems. It typically emerges when the complex structure of the system is simplified during the modeling process, obscuring a part of its underlying behavior. Consequently, the correct description of such a physical model requires knowledge of not only its present, but also its past states, leading to \textit{non-Markovian} dynamics. The resulting interactions depending on the system history, however, introduce a significant layer of mathematical complexity. For this reason, when the memory time is significantly shorter than the time scales characterizing the dynamics, it is typically neglected at the cost of losing information regarding its role in the system behavior. In this dissertation, I introduce a novel methodology, namely the \textit{effective mass approach}, which bridges the Markovian and non-Markovian regimes and provides an accessible framework for analyzing systems with short memory. In this method, the non-Markovian system is approximated with its Markovian counterpart, in which the short-memory effects are encapsulated within an effective mass. I then utilize this approach to demonstrate that the impact of short memory on physical dynamics can be remarkably pronounced. In particular, I show that the directed transport of a Brownian particle in an environment exhibiting temporal correlations can be reversed relative to its behavior in a memoryless medium. Moreover, I propose a setup wherein memory serves as a~control parameter for the generation of random information bits. Collectively, the results of this dissertation underscore the critical importance of short-time correlations in microscopic systems.


\newpage
{
\vspace*{50pt}
\parindent 0pt \normalfont
\centering
\interlinepenalty10000
\huge \scshape Streszczenie\par\nobreak
\vspace*{40pt}
}

\begin{otherlanguage}{polish}
\noindent Pamięć jest nieodłączną cechą dynamiki układów fizycznych. Pojawia się ona zazwyczaj, gdy skomplikowana struktura układu zostaje uproszczona w procesie modelowania, co prowadzi do ukrycia części jego wewnętrznej dynamiki. W konsekwencji opis takich modeli fizycznych wymaga znajomości nie tylko ich stanów teraźniejszych, ale również przeszłych, co prowadzi do zachowania \textit{niemarkowowskiego}. Obecność wynikających z tego oddziaływań zależących od wcześniejszych stanów znacząco komplikuje jednak opis matematyczny takich układów. Z tego powodu, gdy pamięć jest znacznie krótsza niż skale czasowe charakteryzującę dynamikę układu, jest ona zazwyczaj pomijana kosztem utraty informacji dotyczących jej roli w zachowaniu modelu. W niniejszej rozprawie przedstawiam nową metodologię, mianowicie \textit{przybliżenie masy efektywnej}, która stanowi pomost między reżimem markowowskim i niemarkowowskim oraz pozwala na analizę układów z krótką pamięcią. W metodzie tej układ niemarkowowski przybliżany jest jego markowowskim odpowiednikiem, w którym efekty krótkiej pamięci są zawarte w masie efektywnej. Następnie wykorzystuję to podejście, aby wykazać, że wpływ krótkiej pamięci na dynamikę może być bardzo wyraźny. W szczególności pokazuję, że ukierunkowany transport cząstki Browna w środowisku wykazującym korelacje czasowe może ulec odwróceniu względem jej zachowania w ośrodku bez pamięci. Ponadto proponuję układ, w którym pamięć służy jako parametr kontrolny do generowania losowych bitów informacji. Wyniki niniejszej rozprawy podkreślają krytyczne znaczenie korelacji czasowych w układach mikroskopowych.
\end{otherlanguage}

\phantomsection
\chapter*{Acknowledgements}
\addcontentsline{toc}{chapter}{Acknowledgements}

First of all, I would like to thank my supervisor, Jakub Spiechowicz, for introducing me to the world of scientific research and for his constant support throughout the four years of doctoral studies. I am truly grateful for the countless hours of discussion and for his deep commitment to my professional development.

I would also like to thank Karol Bia{\l}as for his companionship at scientific conferences and joint participation in classes in the Doctoral School.

Special thanks go to my family: my parents, Łukasz, and Fred, and all of my friends for their constant support and for always believing in me.

Finally, I am infinitely grateful to You, my dearest Kinga, for your understanding and willingness to help. I could always count on You, and You never let me down.

\phantomsection
\newrefcontext[labelprefix=D]
\defbibnote{publications}{This doctoral dissertation consists of a Guidebook on a collection of the following scientific articles:}
\printbibliography[keyword={phd}, title={List of publications}, prenote=publications]
\addcontentsline{toc}{chapter}{List of publications}

\newrefcontext[labelprefix=A]
\defbibnote{additional}{Additionally, I am the author of the following articles not discussed in this dissertation:}
\printbibliography[keyword=additional, heading=none, prenote=additional]

\chapter*{ }

\hspace{0pt}
\vfill
\begin{center} \huge \scshape
	Guidebook
\end{center}
\vspace*{200pt}
\vfill
\hspace{0pt}
\newpage
\thispagestyle{empty}

}

\mainmatter

\phantomsection
\tableofcontents
\addcontentsline{toc}{chapter}{Contents}

\chapter{Introduction \label{chapter:introduction}}

The correct description of many natural systems requires the knowledge of not only their current, but also their past states. An example is the magnetic hysteresis: the remanent magnetization of a ferromagnet in the absence of a magnetic field depends on its previous magnetization when the field was turned on \cite{Chikazumi1997}.
Another example is the dynamics of cytoplasm, the substance inside biological cells \cite{Bausch1999,Berret2016}. It is a complex environment consisting of a fluidic cytosol and a dynamic polymer network (cytoskeleton) that possesses both viscous and elastic properties \cite{Mogilner2018}, which gives rise to its delayed response to perturbations.
Finally, human reactions to external stimuli may depend on their previous experiences. In this context, what influences human behavior is their \textit{memory} \cite{Tulving2002}, and this term is also used to describe the general dependence of physical systems on their past.

In fact, memory is a universal feature of natural systems \cite{Van_Kampen1998}. Even at the fundamental level of quantum mechanics, a particle coupled to thermal vacuum experiences fluctuations that are correlated in time, i.e., they depend on their previous values \cite{Spiechowicz2021-scirep}. Memory also arises whenever degrees of freedom of the constituents of the system are hidden in the coarse-graining procedure. For example, if we describe the aforementioned ferromagnet only with its total magnetization, we lose information about the orientation of the individual magnetic domains. Such a coarse-grained model consists of much less elements, but in order to predict its behavior, one needs the access to the past values of the magnetic field to which the ferromagnet was exposed. Similarly, in a homogenous picture of cytoplasm, its complexity is hidden and the dynamics of its internal constituents results in its slow relaxation. Since the modeling of physical systems frequently involves projecting out microscopic degrees of freedom, setups with memory are ubiquitous in nature. \hypertarget{target:discoveries}{In recent years the dynamics of such systems has attracted an increased activity and has been studied in the context of active particles \cite{Gomez-Solano2016, Narinder2018, Banerjee2022}, protein folding \cite{Ayaz2021, Dalton2023}, random walk theory \cite{Levernier2022, DAlessandro2021, Guerin2016, Barbier-Chebbah2022}, diffusion \cite{Goychuk2012, Goychuk2022}, stochastic networks \cite{Pelissier2026} and gene expression \cite{Li2025}, to name only a few.}

The emergence of memory is especially pronounced on the mesoscopic level. Mesoscopic objects, such as proteins, constantly exchange energy with their environment, and, despite being much bigger than single molecules, they are small enough to experience the molecular character of their surroundings. The result is their persistent erratic movement known as \textit{Brownian motion} \cite{Einstein1905,Sutherland1905,vonSmoluchowski1906}. The noisy force exerted on the Brownian particles by the environment is frequently modeled with the assumption of its \textit{Markovianity}, i.e., neglecting any temporal correlations. The correlations, however, always arise and may originate from hydrodynamic memory \cite{Franosch2011,Goychuk2019}, activity of the medium \cite{Kanazawa2020,Tucci2022}, or its viscoelasticity \cite{Goychuk2012}. 
The consequence is the inherently \textit{non-Markovian} dynamics of mesoscopic systems. The interactions depending on the past states can significantly influence characteristics of these setups, e.g., the rate of barrier crossing \cite{Ferrer2021,Ginot2022}, and can give rise to phenomena that are absent in corresponding memoryless systems, such as memory-induced Magnus effect \cite{Cao2023}, particle recoil \cite{Chapman2014,Gomez-Solano2015,Pruszczyk2025} and oscillations of driven overdamped particles \cite{Berner2018,Venturelli2023}.

\begin{figure}[tb]
	\centering
	\includegraphics{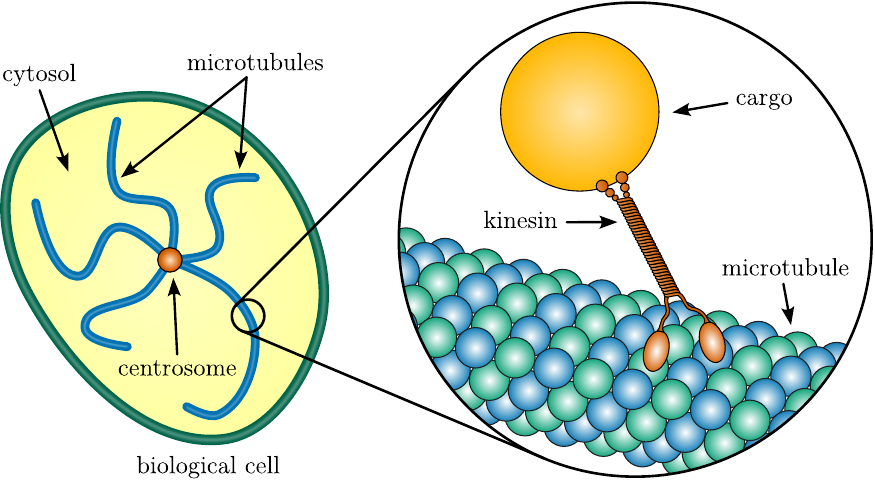}
	\caption{A biological motor operating in a viscoelastic cytoplasm: an example of a non-Markovian system. Microtubules, one type of filaments building up the cytoskeleton, serve as ``tracks'' for the biological motors (such as kinesins) that are responsible for transporting vesicles, organelles and other structures toward or away from the microtubule-organizing centers (such as centrosomes). The non-Markovian character of the motor dynamics results from their interaction with the surrounding intracellular substance (cytoplasm), possessing both viscous and elastic properties, which accounts for its slow relaxation. The illustration is based on the descriptions and diagrams in  Refs.~\cite{Goodson2018,Alberts2002}.}
	\label{fig:cell_kinesin}
\end{figure}

Since viscoelasticity is a general property of numerous soft matter systems, such as polymer networks \cite{Young2011}, micellar solutions \cite{Cates1990}, liquid crystals \cite{Waigh2016} or intracellular environments \cite{Bausch1999}, the dynamics of Brownian particles in environments exhibiting memory has an essential meaning in the physics of proteins, polymers, and cells. A fundamental problem in the study of such systems is the role of memory in \textit{directed transport}. In particular, Brownian particles are the models of biological motors, i.e., proteins responsible for degradation of biomolecules, mitosis, cell differentiation, and particularly for intracellular transport \cite{Julicher1997,Reimann2002,Chowdhury2013} (see Fig.~\ref{fig:cell_kinesin}). Disruptions in the intracellular transport have been linked with various neurodegenerative diseases, such as Alzheimer's and Parkinson's \cite{DeVos2008}. Moreover, some viruses, like HIV, use intracellular transport to infect the host cell's nucleus \cite{Malikov2015,Liu2023}. Our proper understanding of the role of memory in transport at the microscale can thus help in a more effective detection and treatment of various widespread diseases, but it is also crucial for the effective design of artificial micro- and nanomotors, which could serve as agents delivering medicines precisely to the specified point in the human body or could be used to perform mechanical work \cite{Hanggi2009,ErbasCakmak2015}.

\section{Problem formulation and outline}

The non-Markovian character of the dynamics of physical systems introduces a significant layer of mathematical complexity to their description. Consequently, when the memory is much shorter than the characteristic time scales of the system, the history dependence is frequently neglected and the Markovian approximation is applied. Such a treatment, however, hides the effects of short-range temporal correlations, which may significantly alter the system dynamics. In this dissertation, I present a novel methodology for analyzing non-Markovian systems and demonstrate that the effects of short memory may be remarkably profound.

This document is structured as follows. Chapter~\ref{chapter:theory} introduces the fundamental theoretical concepts underlying this work. Specifically, I distinguish between Markovian and non-Markovian stochastic processes, introduce the Generalized Langevin Equation as the primary framework of this study, and detail the Markovian embedding procedure. In Chapter~\ref{chapter:model}, I provide a description of the physical model, define the key observables of interest, outline the numerical methods used to obtain the results, and describe the physical system in which the theoretical results can be verified experimentally.

Chapter~\ref{chapter:effective_mass} presents my first major contribution, namely the \textit{effective mass approach}. This perturbative approximation scheme enables the description of systems with short memory via an equivalent memoryless model, wherein memory effects are recast as a renormalization of the particle's mass. Building on this framework, Chapter~\ref{chapter:transport} demonstrates the profound impact of short memory on the directed transport of Brownian particles. Specifically, I show that the presence of memory can induce a reversal of velocity relative to the memoryless analogue, leading to phenomena such as memory-induced absolute negative mobility and memory-induced current reversal. Additionally, I propose a setup wherein memory serves as a control parameter for generating random bit sequences encoded in the particle's velocity. Finally, Chapter~\ref{chapter:conclusions} provides a summary of the dissertation and offers concluding remarks.

\chapter{Theoretical background \label{chapter:theory}}

In this chapter I introduce the key theoretical concepts appearing in the dissertation. In Sec.~\ref{section:markovian} I formally define the difference between Markovian and non-Markovian stochastic processes. Then, in Sec.~\ref{section:gle} I present the physical model of a Brownian particle in a correlated thermal bath, namely the Generalized Langevin Equation. Finally, I present some classes of memory kernels analyzed in the further part of this work and describe the Markovian embedding technique.

\section{Markovian and non-Markovian processes \label{section:markovian}}

Let us consider a system described by a state vector
\begin{equation}
	\mathbf{y} = 
	\begin{bmatrix}
		y_1 \\ \vdots \\ y_k
	\end{bmatrix} 
\end{equation}
consisting of a finite number of state variables $y_i$, $i\in\{1,\,\dots,\,k\}$. The statistical ensemble of realizations of the trajectory $\mathbf{y}(t)$ is called a stochastic process ($t$ stands for the time variable). At each instant $t^i$ the stochastic process $\mathbf{y}$ is a random variable $\mathbf{y}^i = \mathbf{y}(t^i)$ described by its probability distribution $p^{(1)}\left(\mathbf{y}^i, t^i\right)$.
Similarly, at $n$ time instants the set of random variables $\mathbb{Y}^{(n)} = \{\mathbf{y}^1, \mathbf{y}^2,\dots,\mathbf{y}^n\}$ can be described by a joint probability distribution $p^{(n)}\left(\mathbb{Y}^{(n)}, \mathbf{t}^{(n)}\right)$, where $\mathbf{t}^{(n)} = \{t^1, t^2, \dots, t^n\}$.
Then, the relation
\begin{equation}
	p^{(n+1)}\left(\mathbb{Y}^{(n+1)},\mathbf{t}^{(n+1)}\right) = R^{(n)}\left(\mathbf{y}^{n+1},t^{n+1}\, |\, \mathbb{Y}^{(n)},\mathbf{t}^{(n)}\right)\, p^{(n)}\left(\mathbb{Y}^{(n)}, \mathbf{t}^{(n)}\right)
\end{equation}
defines the transition probability distribution $R^{(n)}\left(\mathbf{y}^{n+1},t^{n+1} \, | \, \mathbb{Y}^{(n)},\mathbf{t}^{(n)}\right)$ to the state $\mathbf{y}^{n+1}$ at time $t^{n+1}$ given that the state vector took values $\mathbf{y}^i \in \mathbb{Y}^{(n)}$ at times $t^i \in \mathbf{t}^{(n)}$.
A stochastic process is \textit{Markovian} if \cite{Hanggi1982}
\begin{equation} \label{eq:Markovian:transprob}
	R^{(n)}\left(\mathbf{y}^{n+1},t^{n+1}\, |\, \mathbb{Y}^{(n)},\mathbf{t}^{(n)}\right) = R^{(1)}\left(\mathbf{y}^{n+1},t^{n+1}\, |\, \mathbf{y}^n,t^n\right).
\end{equation}
From the relation in Eq.~\eqref{eq:Markovian:transprob} it follows that for a Markovian process the transition probability distribution depends only on the latest value of the state vector $\mathbf{y}$ (it does not depend on the past values $\mathbf{y}^i$ at times $t^i < t^n$). The Markovian process is thus fully described by its initial probability distribution $p^{(1)}\left(\mathbf{y}^1,t^1\right)$ and the one-step transition probability distribution $R^{(1)}\left(\mathbf{y}^{n+1},t^{n+1}\, |\, \mathbf{y}^n,t^n\right)$ obeying the Chapman-Kolmogorov equation
\begin{multline} \label{eq:ChapKol}
	R^{(1)}\left(\mathbf{y}^{n+1},t^{n+1}\, |\, \mathbf{y}^{n-1},t^{n-1}\right) =\\
	= \int\!\cdots\!\int_{\mathbb{D}} R^{(1)}\left(\mathbf{y}^{n+1},t^{n+1}\, |\, \mathbf{y}^n,t^n\right)\, R^{(1)}\left(\mathbf{y}^n,t^n\, |\, \mathbf{y}^{n-1},t^{n-1}\right)\, \mathrm{d}\mathbf{y}^n,
\end{multline}
for fixed $t^{n-1} < t^n < t^{n+1}$ (integration is over the $k$-dimensional domain $\mathbb{D}$ of the state vector $\mathbf{y}$). For a \textit{non-Markovian} process the transition probability distribution depends on the previous values of the state vector $\mathbf{y}$ (the process has memory of its previous states) and in general it does not obey the Chapman-Kolmogorov equation \eqref{eq:ChapKol}.

\section{Microscopic model of Brownian motion \label{section:gle}}

In this section I introduce the main theoretical framework on which my work is based, namely the Generalized Langevin Equation. I start with presenting the memoryless Langevin Equation for a Brownian particle in an uncorrelated bath. Then I derive its generalized form taking into account the internal degrees of freedom of the environment.

\subsection{Langevin Equation}

One of the earliest mathematical models of Brownian motion dates back to 1908 and was proposed by Paul Langevin \cite{Lemons1997}. In this model, the one-dimensional dynamics of a Brownian particle with mass $m$ and position $x$ is described with a Newton's equation of motion
\begin{equation} \label{eq:LE}
	\tag{LE}
	m\dot{v}(t) + \gamma v(t) = F(x, t) + \xi(t),
\end{equation}
where $v(t) = \dot{x}(t)$ is the particle's velocity, $F(x, t)$ represents arbitrary external forces acting on the particle and dot indicates differentiation with respect to time $t$. The particle is driven by the noisy force $\xi(t)$ exerted by the environment and resulting from the collisions of the surrounding molecules with the particle. At the same time, the particle dissipates its energy to the bath via the Stokes frictional force $-\gamma v(t)$, where $\gamma=6\pi\nu a$ is the friction coefficient, $\nu$ is the viscosity of the surrounding medium and $a$ is the particle's radius. In this treatment the force $\xi(t)$ responsible for the erratic motion of the particle is assumed to be a Gaussian process with the following properties%
\begin{subequations}
\begin{align}
	\langle \xi(t) \rangle &= 0, \label{eq:xi0} \\
	\langle \xi(t) \xi(t') \rangle &= 2\gamma k_\mathrm{B} T \delta(t-t'), \label{eq:xitxitp}%
\end{align}
\end{subequations}
where $k_\mathrm{B}$ is the Boltzmann constant, $T$ is the temperature of the bath, $\delta(t)$ is the Dirac's delta, and the triangular brackets indicate averaging over an ensemble of realizations of the random force $\xi(t)$. The assumption of Gaussianity of $\xi(t)$ is justified by the large number of collisions of the Brownian particle with the surrounding molecules and follows from the central limit theorem. The condition of its vanishing mean in Eq.~\eqref{eq:xi0} is required for thermal equilibrium. Finally, Eq.~\eqref{eq:xitxitp} is a manifestation of the fluctuation-dissipation theorem interconnecting autocorrelation function of the random force (or equivalently its spectrum) with the Stokes friction experienced by the particle \cite{Kubo1966}. 
In particular, Eq.~\eqref{eq:xitxitp} states that the subsequent values of the force $\xi(t)$ are not correlated.

The particle's state at a given time instant is unambiguously described with a state vector
\begin{equation}
	\mathbf{y}(t) = \begin{bmatrix}
		x(t) \\ v(t)
	\end{bmatrix}.
\end{equation}
According to Eq.~\eqref{eq:LE}, the time evolution of $\mathbf{y}(t)$ is determined by the following equation
\begin{equation} \label{eq:LE:statevec:evol}
	\dot{\mathbf{y}}(t) = 
	\begin{bmatrix}
		\dot{x}(t) \\ \dot{v}(t)
	\end{bmatrix} = 
	\begin{bmatrix}
		v(t) \\
		\frac{1}{m}\left[ -\gamma v(t) + F(x, t) + \xi(t)\right]
	\end{bmatrix}.
\end{equation}
The right-hand side of Eq.~\eqref{eq:LE:statevec:evol} depends only on the current value of the state vector $\mathbf{y}(t)$ (not on its past values) and on the random force $\xi(t)$, whose values at subsequent time instants are independent. Consequently, the ensemble of trajectories $\mathbf{y}(t)$ for different initial positions $x(0)$, velocities $v(0)$ and realizations of the thermal noise $\xi(t)$ form a \textit{Markovian} process.

\subsection{Generalized Langevin Equation}

The assumption of the Markovian character of the Brownian particle dynamics is correct only approximately. In actual systems the collision times of the molecules are never instantaneous. The constituents of the bath may have non-zero relaxation times, which results in a correlated random force experienced by the particle. The generalized form of Eq.~\eqref{eq:LE} can be derived starting with a model consisting of a Brownian particle coupled bilinearly with a bath of harmonic oscillators with masses $m_i$ and frequencies $\omega_i$ \cite{Zwanzig1973,Hanggi1997}. The total Hamiltonian of the system then reads
\begin{equation} \label{eq:H}
	H = \frac{1}{2}m v^2 + U(x, t) + \sum\limits_{i}\left[ \frac{1}{2}m_i v_i^2 + \frac{m_i \omega_i^2}{2} \left( x_i - \frac{c_i}{m_i\omega_i^2}x \right)^2 \right],
\end{equation}
where the potential $U(x, t)$ is defined by the relation
\begin{equation}
	F(x, t) = -\frac{\partial U(x, t)}{\partial x},
\end{equation}
the indexed variables $x_i$ and $v_i$ represent the positions and velocities of the bath oscillators, and the constants $c_i$ describe the coupling between the Brownian particle and the bath.
The equation of motion for the Brownian particle then reads
\begin{equation} \label{eq:GLE:y}
	\dot{\mathbf{y}}(t) = 
	\begin{bmatrix}
		v(t) \\
		\frac{1}{m} F(x, t) + \sum\limits_{i} \frac{c_i}{m}\left( x_i(t) - \frac{c_i}{m_i\omega_i^2}x(t) \right)
	\end{bmatrix},
\end{equation}
and for the bath oscillators
\begin{equation} \label{eq:GLE:yi}
	\dot{\mathbf{y}}_i(t) = 
	\begin{bmatrix}
		\dot{x}_i(t) \\ \dot{v}_i(t)
	\end{bmatrix} = 
	\begin{bmatrix}
		v_i(t) \\
		-\omega_i^2 x_i + \frac{c_i}{m_i} x
	\end{bmatrix}.
\end{equation}
Solving Eq.~\eqref{eq:GLE:yi} for the positions $x_i(t)$ of the bath particles and inserting the results into Eq.~\eqref{eq:GLE:y} gives
\begin{equation} \label{eq:GLE}
	\tag{GLE}
\boxed{
	m\dot{v}(t) + \gamma\int_0^t K(t-t') v(t')\mathrm{d}t' = F(x, t) + \eta(t),
}
\end{equation}
where
\begin{equation} \label{eq:K}
	K(t) = \frac{1}{\gamma}\sum\limits_{i} \frac{c_i^2}{m_i\omega_i^2} \cos(\omega_i t)
\end{equation}
is a \textit{memory kernel}, and
\begin{equation} \label{eq:eta}
	\eta(t) = \sum\limits_{i} c_i\left[ \left( x_i(0) - \frac{c_i}{m_i\omega_i^2}x(0) \right)\cos(\omega_i t) + \frac{v_i(0)}{\omega_i}\sin(\omega_i t) \right]
\end{equation}
is a correlated random force with zero-mean obeying the second \textit{fluctuation-dissipation relation} \cite{Kubo1966}
\begin{equation} \label{eq:FDrel}
\boxed{
	\langle \eta(t) \eta(t') \rangle = \gamma k_\mathrm{B} T K(t-t').
}
\end{equation}
The definition of the force $\eta(t)$ does not consist of any stochastic component, however, assuming a large number of the bath oscillators with unknown initial positions $x_i(0)$ and velocities $v_i(0)$, it can be effectively treated as a random force, in the same manner as the uncorrelated random force $\xi(t)$ approximates the deterministic force resulting from the collisions with the surrounding molecules in the memoryless Langevin Equation \eqref{eq:LE}. Moreover, if the bath particles form a Gibbs equilibrium ensemble, the force $\eta(t)$ is a Gaussian process with vanishing mean, i.e., $\langle \eta(t) \rangle = 0$, and the relation in Eq.~\eqref{eq:FDrel} is fulfilled \cite{Hanggi1997}. Eq.~\eqref{eq:GLE} is the generalized version of Eq.~\eqref{eq:LE} taking into account the internal degrees of freedom of the bath and so is called the \textit{Generalized Langevin Equation}.

In analogy with Eq.~\eqref{eq:LE:statevec:evol}, the time evolution of the state vector $\mathbf{y}(t)$ is determined by the following equation
\begin{equation} \label{eq:GLE:statevec:evol}
	\dot{\mathbf{y}}(t) = 
	\begin{bmatrix}
		v(t) \\
		\frac{1}{m}\left[ -\gamma\int_0^t K(t-t') v(t')\mathrm{d}t' + F(x, t) + \eta(t) \right]
	\end{bmatrix}.
\end{equation}
The right-hand side of Eq.~\eqref{eq:GLE:statevec:evol} depends on all the values of the particle velocity $v(t)$ from 0 to $t$. Moreover, from Eq.~\eqref{eq:FDrel} it follows that the subsequent values of the random force $\eta(t)$ are not independent. This means that in the Generalized Langevin Equation formalism the ensemble of the trajectories $\mathbf{y}(t)$ generally form a \textit{non-Markovian} process.

\subsection{Memory kernels \label{sec:theory:kernels}}

The complete information about the character of the particle-bath interaction is contained in the memory kernel $K(t)$. From Eq.~\eqref{eq:K} it follows that $K(t)$ is a real-valued even function, i.e.
\begin{equation}
	K(-t) = K(t) \in \mathbb{R}.
\end{equation}
Since $K(t)$ is proportional to the autocorrelation function of the random force $\eta(t)$ [see the fluctuation-dissipation relation in Eq.~\eqref{eq:FDrel}] it must be positive semi-definite \cite{Berne1970}. In particular it means that
\begin{equation}
	K(0) \geq |K(t)|,
\end{equation}
so that its value at $t=0$ is positive and is a global maximum. Moreover, the one-sided Fourier transform of $K(t)$ defined as
\begin{equation}
	\tilde{K}(\omega) = \int_0^\infty e^{-\mathrm{i}\omega t} K(t) \mathrm{d} t = \tilde{K}_\mathrm{R}(\omega) + \mathrm{i}\tilde{K}_\mathrm{I}(\omega)
\end{equation}
is analytic in the lower half of the complex plane and vanishes as $|\omega|\to\infty$, thus its real and imaginary parts obey the Kronig-Kramers relations
\begin{align}
	\tilde{K}_\mathrm{R}(\omega) &= \text{p.v.}\left[\frac{1}{\pi} \int_{-\infty}^{\infty} \frac{\tilde{K}_\mathrm{I}(\omega')}{\omega' - \omega} \mathrm{d}\omega'\right], \\
	\tilde{K}_\mathrm{I}(\omega) &= \text{p.v.}\left[-\frac{1}{\pi} \int_{-\infty}^{\infty} \frac{\tilde{K}_\mathrm{R}(\omega')}{\omega' - \omega} \mathrm{d}\omega'\right],
\end{align}
where ``p.v.'' denotes the Cauchy principal value. $\tilde{K}_\mathrm{R}(\omega)$ is also proportional to the power spectrum of $\eta(t)$ and thus it must be non-negative \cite{Berne1970}.

A reasonable assumption about the memory kernel states that it vanishes as $t\to\infty$, i.e.
\begin{equation}
	\lim\limits_{t\to\infty} K(t) = 0.
\end{equation}
Intuitively, it means that after a long time the influence of the past states on the current dynamics is negligible. If the memory kernel does not vanish as $t\to\infty$, the particle becomes trapped \cite{Goychuk2012}. A stronger assumption for the memory kernel requires it to be integrable
\begin{equation}
	\int_0^\infty K(t) \mathrm{d}t < \infty,
\end{equation}
which ensures that the Brownian particle dragged with a constant velocity will asymptotically experience constant and finite friction. This assumption is sometimes removed if only transient times are considered.

\begin{figure}[tb]
	\centering
	\includegraphics{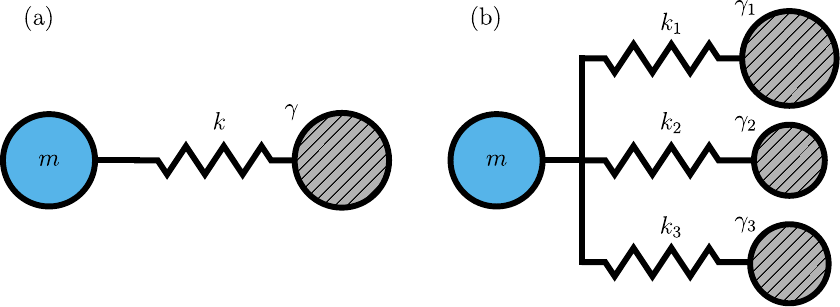}
	\caption{Schematic representation of two models of viscoelasticity. (a) In a Maxwell model the Brownian particle (blue, solid) is connected to a fictitious overdamped bath particle (gray, hatched) with a damping coefficient $\gamma$ through a spring with elasticity constant $k=\gamma/\tau$ (zigzag line). (b) In the generalized Maxwell model the bath is modeled with $n>1$ overdamped particles characterized by different friction coefficients $\gamma_i = \alpha_i\gamma$ and spring constants $k_i = \gamma_i/\tau_i$. Here the case $n=3$ is visualized.}
	\label{fig:Maxwell_model}
\end{figure}

There is a plethora of particular forms of the memory kernel appearing in the literature. Here I present only a few classes of them that will appear in the later part of this work. First, the Generalized Langevin Equation \eqref{eq:GLE} reduces to the memoryless Langevin Equation \eqref{eq:LE} for
\begin{equation} \label{eq:K0}
	K_0(t) = 2\delta(t),
\end{equation}
where $\delta(t)$ is the Dirac's delta, and I follow the convention $\int_0^t \delta(t')\mathrm{d}t' = 1/2$ consistent with the representation of the Dirac's delta as a limit of a sequence of even functions. The ``memory'' kernel $K_0(t)$ in fact represent the instantaneous (memoryless) interaction between the bath and the particle. It is a suitable choice if the memory time of the bath is so short that no memory effects are observed at the considered time scale. One of the most common models of $K(t)$ that actually captures the non-Markovian effects is the exponential kernel
\begin{equation} \label{eq:KM}
\boxed{
	K_\mathrm{M}(t) = \frac{1}{\tau}e^{-|t|/\tau},
}
\end{equation}
where $\tau$ is the memory (correlation) time. The form of $K_\mathrm{M}(t)$ follows from the model of viscoelasticity proposed in 1867 by James Clerk Maxwell \cite{Maxwell1867} (well before the derivation of the Generalized Langevin Equation). Such a memory kernel also appears when the Brownian particle is connected to a fictitious overdamped (intertialess) bath particle by a spring with an elastic constant $\gamma/\tau$ \cite{Goychuk2012,Ginot2022} [see Fig.~\ref{fig:Maxwell_model}(a)].
The memory kernel $K_\mathrm{M}(t)$ is characterized by a single relaxation time $\tau$. More complex environments may exhibit a whole spectrum of relaxation times $\tau_i$ and the memory kernel can be composed as a weighted sum of $n$ exponentially decaying functions, i.e.
\begin{equation} \label{eq:KGM}
	K_\mathrm{GM} = \sum\limits_{i=1}^n \frac{\alpha_i}{\tau_i} e^{-|t|/\tau_i},
\end{equation}
where $\alpha_i$ specify the contribution of each of the sum elements to the kernel. In this generalized Maxwell model the bath is modeled with $n$ fictitious overdamped particles characterized by different friction coefficients and connected with the Brownian particle by springs with different elastic constants [see Fig.~\ref{fig:Maxwell_model}(b)]. This model can accurately represent the bath on its own \cite{Ginot2022-eigen}, but it can also serve as an approximation of other kernel types, such as the power-law $K_\mathrm{P}(t) \propto |t|^{-\beta}$ \cite{Goychuk2009,Beylkin2005} or stretched exponential $K_\mathrm{S}(t) \propto \exp(-|t|^\beta)$ \cite{Palmer1984} kernels.

Additionally, if the response of the bath is characterized by clearly separated time scales, the memory kernel may consist of both $K_0(t)$ representing the fast (instantaneous) response of the medium, and $K_\mathrm{M}(t)$ or $K_\mathrm{GM}(t)$ describing the delayed reaction.

\subsection{Markovian embedding \label{subsection:embedding}}

As stated before, in the Generalized Langevin Equation formalism the process $\mathbf{y}(t)$ describing the particle dynamics is generally non-Markovian [with exception to the memory kernel $K_0(t)$, see Eq.~\eqref{eq:K0}]. There is, however, a class of memory kernels for which the state vector $\mathbf{y}$ can be extended with $n$ auxiliary variables $u_i(t)$ describing the dynamics of the thermal bath to form a Markovian process \cite{Siegle2010}. In particular, let us consider a set of equations
\begin{subequations} \label{eq:embedding:set}
\begin{align}
	\dot{x}(t) &= v(t), \\
	m\dot{v}(t) &= F(x, t) + \mathbf{g}^\mathrm{\mathbf{T}} \mathbf{u}(t), \label{eq:embedding:set:v} \\
	\dot{\mathbf{u}}(t) &= -\gamma v(t)\mathbf{h} - \mathbb{A}\mathbf{u}(t) + \mathbb{C}\boldsymbol{\xi}(t), \label{eq:embedding:set:u}
\end{align}
\end{subequations}
where $\mathbf{u}(t)$ is the vector of auxiliary variables $u_i(t)$, $\mathbf{g}$ and $\mathbf{h}$ are constant vectors with $n$ elements, $\mathbb{A}$ and $\mathbb{C}$ are constant $n\times n$ matrices, $\boldsymbol{\xi}(t)$ is an $n$-element vector of uncorrelated Gaussian white noises satisfying $\langle \xi_i(t) \xi_j(t') \rangle = 2\gamma k_B T\delta_{ij}\delta(t-t')$ (where $\delta_{ij}$ is the Kronecker delta), and the upper index $\mathrm{\mathbf{T}}$ denotes a transpose of a vector or a matrix. The evolution equation for the extended process
\begin{equation}
	\mathbf{y}_\mathrm{E}(t) = 
	\begin{bmatrix}
		x(t) \\ v(t) \\ u_1(t) \\ \vdots \\ u_n(t)
	\end{bmatrix}
\end{equation}
depends only on the current values of $\mathbf{y}_\mathrm{E}(t)$ and contains uncorrelated noisy forces $\xi_i(t)$, thus the extended process is Markovian.

It is then possible to integrate Eq.~\eqref{eq:embedding:set:u} for the auxiliary variables $u_i(t)$. After substituting it to Eq.~\eqref{eq:embedding:set:v} the result has a form of a Generalized Langevin Equation \eqref{eq:GLE} with the following memory kernel
\begin{equation} \label{eq:embedding:K}
	K(t) = \mathbf{g}^\mathrm{\mathbf{T}} e^{-\mathbb{A}t} \mathbf{h}.
\end{equation}
In order for the model to be consistent with the fluctuation-dissipation relation \eqref{eq:FDrel}, the following relation must hold
\begin{equation} \label{eq:embedding:Gg}
	\mathbb{G}\mathbf{g} = \mathbf{h},
\end{equation}
where the $n\times n$ matrix $\mathbb{G}$ is defined by the equation
\begin{equation} \label{eq:embedding:CC}
	\mathbb{C}\mathbb{C}^\mathrm{\mathbf{T}} = \frac{1}{2} \left( \mathbb{A}\mathbb{G} + \mathbb{G}\mathbb{A}^\mathrm{\mathbf{T}}\right).
\end{equation}
Moreover, the vector $\mathbf{u}(0)$ must be Gaussian distributed with the covariance matrix given by $\gamma k_\mathrm{B} T\mathbb{G}$. 
The Generalized Langevin Equation \eqref{eq:GLE} describing the non-Markovian process $\mathbf{y}(t)$ is thus equivalent to a set of equations \eqref{eq:embedding:set} defining the extended Markovian process $\mathbf{y}_\mathrm{E}(t)$ if only the memory kernel is representable as in Eq.~\eqref{eq:embedding:K} and obeys the equations~\eqref{eq:embedding:Gg} and \eqref{eq:embedding:CC}.
The procedure of extending the state vector corresponding to a non-Markovian process with auxiliary variables to obtain a Markovian model is called \textit{Markovian embedding}.

The memory kernel in Eq.~\eqref{eq:embedding:K} can be represented as a sum of the functions $t^k \exp(-\lambda_i t)$, where $\lambda_i$ are unique eigenvalues of the matrix $\mathbb{A}$, and $k \in \{0,\dots, m_i-1\}$, where $m_i$ is the algebraic multiplicity of the $i$th unique eigenvalue. If all $m_i = 1$ and $\lambda_i \in \mathbb{R}$, the kernel is represented by a sum of exponential decays and corresponds to $K_\mathrm{GM}(t)$ appearing in the generalized Maxwell model of viscoelasticity [see Eq.~\eqref{eq:KGM}]. If some of the eigenvalues $\lambda_i$ are complex, the kernel may exhibit decaying oscillations, as it is e.g.~in a model of harmonic noise \cite{SchimanskyGeier1990}.

In the simplest non-trivial case $n=1$ all of the vectors and matrices appearing in Eq.~\eqref{eq:embedding:set} reduce to constants. Setting 
\begin{equation}
	\mathbf{g} = 1,\quad \mathbf{h} = \mathbb{A} = \mathbb{C} = \mathbb{G} = \frac{1}{\tau}
\end{equation}
produces an exponentially decaying kernel $K_\mathrm{M}(t)$, the same as in the Maxwell's model of viscoelasticity [see Eq.~\eqref{eq:KM}]. The set of equations describing the joint particle and bath dynamics then reads
\begin{subequations} \label{eq:embedding:set:Maxwell}
\begin{align}
	\dot{x}(t) &= v(t), \\
	m\dot{v}(t) &= F(x, t) + u(t), \\
	\tau \dot{u}(t) &= -\gamma v(t) - u(t) + \xi(t).
\end{align}
\end{subequations}
The memory kernel $K_\mathrm{GM}(t)$ appearing in the generalized Maxwell model of viscoelasticity can be obtained by setting
\begin{equation}
	\mathbf{g} = \begin{bmatrix}
		1 \\
		\vdots \\
		1
	\end{bmatrix}\!,\;\,
	\mathbf{h} = \begin{bmatrix}
		\frac{\alpha_1}{\tau_1} \\
		\vdots \\
		\frac{\alpha_n}{\tau_n}
	\end{bmatrix}\!,\;\
	\mathbb{A} = 
	\begin{bNiceMatrix}[nullify-dots]
		\frac{1}{\tau_1}	  & &  \Block[l]{1-1}{\raisebox{-2.0ex}{\hspace{-0.5ex}\LARGE 0}}\\
							  & \Ddots &				\\
		\Block[r]{1-1}{\raisebox{1.0ex}{\hspace{0.5ex}\LARGE 0}} & 					  & \frac{1}{\tau_n}
	\end{bNiceMatrix},\;
	\mathbb{C} = 
	\begin{bNiceMatrix}[nullify-dots]
		\frac{\sqrt{\alpha_1}}{\tau_1}	  & &  \Block[l]{1-1}{\raisebox{-2.0ex}{\hspace{-0.0ex}\LARGE 0}}\\
		& \Ddots &				\\
		\Block[r]{1-1}{\raisebox{1.0ex}{\hspace{0.0ex}\LARGE 0}} & 					  & \frac{\sqrt{\alpha_n}}{\tau_n}
	\end{bNiceMatrix},\;
	\mathbb{G} = 
	\begin{bNiceMatrix}[nullify-dots]
		\frac{\alpha_1}{\tau_1}	  & &  \Block[l]{1-1}{\raisebox{-2.0ex}{\hspace{-0.5ex}\LARGE 0}}\\
		& \Ddots &				\\
		\Block[r]{1-1}{\raisebox{1.0ex}{\hspace{0.5ex}\LARGE 0}} & 					  & \frac{\alpha_n}{\tau_n}
	\end{bNiceMatrix}.
\end{equation}
The corresponding set of equations of motion reads
\begin{subequations} \label{eq:embedding:set:gMaxwell}
	\begin{align}
		\dot{x}(t) &= v(t), \\
		m\dot{v}(t) &= F(x, t) + \sum\limits_{i=1}^n u_i(t), \\
		\tau_i \dot{u}_i(t) &= -\alpha_i\gamma v(t) - u_i(t) + \sqrt{\alpha_i}\xi_i(t).
	\end{align}
\end{subequations}

Summarizing, the Markovian embedding procedure maps the non-Markovian dynamics of the particle described with a state vector $\mathbf{y}$ to an extended phase space containing the extended state vector $\mathbf{y}_\mathrm{E}$, for which the joint particle-bath dynamics is Markovian. Conversely, the dynamics of the state vector $\mathbf{y}$ described with the Generalized Langevin Equation \eqref{eq:GLE} is non-Markovian, because the bath degrees of freedom are hidden in the model. The Markovian embedding procedure is especially useful in numerical analysis of Eq.~\eqref{eq:GLE}, since it allows for utilization of the algorithms developed for Markovian systems.

\chapter{Model and methods\label{chapter:model}}

In this chapter I present the details of the model analyzed in this dissertation. First, I specify the external force acting on the particle and the type of memory kernel describing the thermal bath. Then, I define the quantities of interest appearing in the further part of this work. Next, I recast the Generalized Langevin Equation to a dimensionless form and describe the computational methods implemented to solve the equation of interest numerically. Finally, I discuss the possible experimental realizations of the presented model.

\section{Details of the model}

\begin{figure}[tb]
	\centering
	\includegraphics{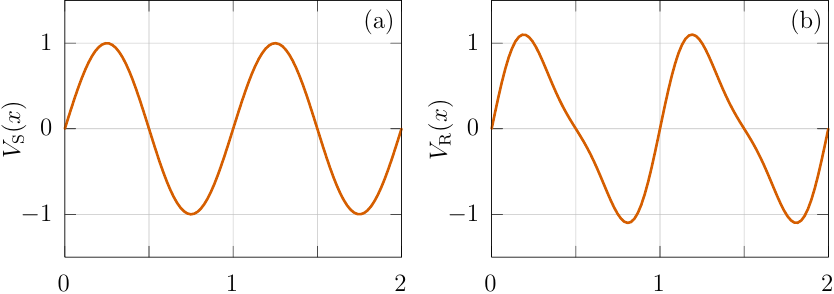}
	\caption{Two classes of periodic potentials appearing in this work: (a) a symmetric sinusoidal potential $V_\mathrm{S}(x)$; (b) an asymmetric ratchet potential $V_\mathrm{R}(x)$.}
	\label{fig:potentials}
\end{figure}

In the following part of this dissertation I focus on the dynamics of a Brownian particle in a periodic potential $V(x)$ reading either
\begin{equation} \label{eq:Vs}
	V_\mathrm{S}(x) = \Delta V \sin(2\pi x/L)
\end{equation}
and representing a \textit{symmetric} landscape, or
\begin{equation} \label{eq:Vr}
	V_{\mathrm{R}}(x) = \Delta V \left[\sin(2\pi x/L) + \frac{1}{4}\sin(4\pi x/L)\right]
\end{equation}
called a \textit{ratchet} potential, whose slopes are \textit{asymmetric} \cite{Hanggi2009} (see Fig.~\ref{fig:potentials}). The periodic potential may represent, e.g., a substrate on which the microscopic object modeled as a Brownian particle resides, or a potential field generated with optical tweezers.

Moreover, I consider a particle exposed to two external forces: a constant bias $f$ and a periodic driving $a \cos(\omega t)$. The time-periodic force ensures that the particle is not in thermal equilibrium with the surrounding bath (as it frequently happens in e.g.~biological systems), whereas the constant bias breaks the spatial symmetry and is necessary for the observation of directed transport in the symmetric potential $V_\mathrm{S}(x)$. The total external force acting on the particle then reads
\begin{equation} \label{eq:Fxt}
	F(x, t) = -\frac{\partial V(x)}{\partial x} + a\cos(\omega t) + f.
\end{equation}

Finally, I consider the exponentially decaying memory kernel $K_\mathrm{M}(t)$ characterized by a single relaxation time $\tau$, see Eq.~\eqref{eq:KM}.

\section{Quantities of interest}

In the following part of this work, the main quantity of interest will be the \textit{average velocity} measuring the net directed transport of the ensemble of particles
\begin{equation}
	\langle v \rangle = \lim\limits_{t\to\infty} \frac{1}{t}\int_0^t \langle \dot{x}(t') \rangle\mathrm{d}t',
\end{equation}
where the triangular brackets indicate the average over an ensemble of particles starting with different initial positions $x(0)$ and velocities $v(0)$ and exposed to different realizations of the thermal noise $\eta(t)$. In the presence of the driving force $a\cos(\omega t)$, one can also define velocity averaged over the period $\mathsf{T} = 2\pi/\omega$ of the driving 
\begin{equation} \label{eq:vperiod}
	\mathsf{v}(t) = \frac{1}{\mathsf{T}} \int_t^{t+\mathsf{T}} v(t')\mathrm{d}t'.
\end{equation}
If the instantaneous velocity $v(t)$ is periodic with the period $\mathsf{T}$, this \textit{period-averaged velocity} eliminates its oscillations that may obscure the underlying non-periodic dynamics and reaches a stationary state in the long-time limit.

Another quantity characterizing directed transport is the \textit{mobility} $\mu(f)$, which relates the particle's average velocity $\langle v \rangle$ to the constant bias $f$ breaking the spatial symmetry, namely
\begin{equation}
	\mu(f) = \frac{\langle v \rangle(f)}{f}.
\end{equation}
In the small bias limit $f\to 0$ one expects the linear response of the system described by the \textit{absolute mobility}
\begin{equation} \label{eq:mu0}
	\mu_0 = \lim\limits_{f\to0} \mu(f).
\end{equation}

\section{Dimensionless form of GLE \label{section:dimensionless}}

The Generalized Langevin Equation \eqref{eq:GLE} can be recast to a dimensionless form. A proper choice of time and length units allows for elimination of several parameters describing the model, thus reducing the effective parameter space. Moreover, in the dimensionless equation only relative values of the physical quantities matter, so the results are independent of the particular physical system described by the mathematical model.

With the external force chosen as in Eq.~\eqref{eq:Fxt}, the most natural length unit is the period of the potential $L$. For the time unit $\tau_0$, there are various reasonable choices corresponding to different characteristic times associated with the particle dynamics (see e.g.~Ref.~\cite{Wisniewski2022-ANM}). In this dissertation I set
\begin{equation}
	\tau_0 = \frac{\gamma L^2}{\Delta V},
\end{equation}
which is related to the time scale of the relaxation of an overdamped (inertialess) particle in the periodic potential $V(x)$ \cite{Machura2008}. After multiplying Eq.~\eqref{eq:GLE} by $\tau_0/(\gamma L)$, the position $x$ and time $t$ are replaced by their dimensionless counterparts $\hat{x} = x/L$ and $\hat{t} = t/\tau_0$. The friction coefficient $\gamma$, magnitude of the potential $\Delta V$ and its period $L$ cancel out to unity, and the resulting equation for dimensionless variables reads
\begin{equation} \label{eq:GLE:dimless}
	\hat{m}\dot{\hat{v}}(\hat{t}) + \int_0^{\hat{t}} \hat{K}(\hat{t}-\hat{t}')\hat{v}(\hat{t}')\mathrm{d}\hat{t}' = \hat{F}(\hat{x},\hat{t}) + \hat{\eta}(\hat{t}),
\end{equation}
where the dimensionless mass reads
\begin{equation}
	\hat{m} = \frac{m}{\gamma\tau_0},
\end{equation}
and the external force scales to
\begin{equation}
	\hat{F}(\hat{x},\hat{t}) = -\frac{\partial \hat{V}(\hat{x})}{\partial \hat{x}} + \hat{a}\cos(\hat{\omega}\hat{t}) + \hat{f},
\end{equation}
where
\begin{equation}
	\hat{a} = a\frac{L}{\Delta V},\quad \hat{f} = f\frac{L}{\Delta V},\quad \hat{\omega} = \omega\tau_0.
\end{equation}
The dimensionless periodic potentials read
\begin{equation}
	\hat{V}_\mathrm{S}(\hat{x}) = \sin(2\pi\hat{x}),\quad \hat{V}_\mathrm{R}(\hat{x}) = \sin(2\pi\hat{x})+\frac{1}{4}\sin(4\pi\hat{x}),
\end{equation}
and the memory kernel scales to
\begin{equation}
	\hat{K}(\hat{t}) = \tau_0 K(t).
\end{equation}
The rescaled thermal noise force obeys the fluctuation-dissipation relation in the form
\begin{equation}
	\langle \hat{\eta}(\hat{t}) \hat{\eta}(\hat{t}') \rangle = \theta \hat{K}(\hat{t}-\hat{t}'),
\end{equation}
with
\begin{equation}
	\theta = \frac{k_\mathrm{B} T}{\Delta V}
\end{equation}
being the dimensionless temperature, relating the thermal energy $k_\mathrm{B}T$ to the height of the potential barrier determined by $\Delta V$.

\section{Numerical simulations \label{section:simulations}}

The Generalized Langevin Equation is a second-order stochastic integro-differential equation. Due to the nonlinearity (in the particle's position $x$) of the potential $V(x)$, its analytical solution is unattainable. In this section I describe the methods applied to solve the equation of interest numerically.

As stated in subsection \ref{subsection:embedding}, Eq.~\eqref{eq:GLE} with a memory kernel $K(t)$ representable as in Eq.~\eqref{eq:embedding:K} can be recast to a set of first-order stochastic differential equations \eqref{eq:embedding:set}. For such a class of models there exist well-established numerical algorithms for solving the equations of motion numerically. In this work a weak second-order predictor-corrector scheme was implemented for this task \cite{Platen2010}. The time step of the simulations was set to $\Delta t \in \left(\frac{\mathsf{T}}{4000},\ \frac{\mathsf{T}}{200}\right)$, where $\mathsf{T} = 2\pi/\omega$ is the fundamental period of the external driving $a \cos(\omega t)$. The ensemble averages were calculated over $2^{12}$-$2^{18}$ system trajectories starting with different initial conditions and driven by different realizations of the random force $\xi(t)$. All the simulation parameter sets were verified for their numerical convergence.

Since all of the trajectories in the ensemble are described by the same equations of motion, the numerical calculations were performed in parallel on modern desktop Graphics Processing Unit (GPU) by use of the CUDA environment \cite{CUDA}. This method allowed for a significant speed-up of the calculations, reducing the computational time by up to three orders of magnitude compared to the Computer Processing Unit (CPU) methods \cite{Spiechowicz2015}.

\section{Experimental verification}

The theoretical predictions presented in this dissertation can be corroborated experimentally in a system consisting of a colloid driven with an optical trap in a viscoelastic medium. Typically, the Brownian particle is a silica probe with a diameter in the order of micrometers, and the viscoelastic medium is composed of wormlike micelles or a polymer solution \cite{Chapman2014,Jain2021,Ginot2022-eigen}. The periodic potential and the external force can be implemented with a rotating circular array of optical traps \cite{Evstigneev2008} or can be simulated with optical tweezers using a high-precision feedback protocol \cite{Albay2018}. Such a setup acts as a prototypical experimental platform for investigating the fundamental problems arising in the field of non-Markovian dynamics and has already been used to study the rate of barrier crossing \cite{Ferrer2021,Ginot2022}, giant diffusion \cite{Evstigneev2008}, stochastic resonance \cite{Babic2004}, energy recuperation \cite{Ginot2025} and optimal control \cite{Loos2024}. The experimental methods nowadays offer unprecedented accuracy, which allowed for, e.g., tracking the Brownian particles at time scales shorter than their relaxation times \cite{Kheifets2014, Madsen2021} and observation of their ballistic motion \cite{Huang2011,Hammond2017,Boynewicz2026}. Such precise measurements allow for verifying various theoretical predictions, such as the Maxwell-Boltzmann distribution \cite{Li2010,Mo2015}, and can be also applied to test the results presented in this dissertation.

\chapter{Effective mass approach \label{chapter:effective_mass}}

The Langevin Equation \eqref{eq:LE} is a memoryless limit of the Generalized Langevin Equation \eqref{eq:GLE}. 
For example, the memory kernel appearing in the Maxwell's model of viscoelasticity $K_\mathrm{M}(t)$ [see Eq.~\eqref{eq:KM}] converges 
to the kernel $K_0(t)$ [see Eq.~\eqref{eq:K0}] if the memory time $\tau \to 0$. Consequently, when all the memory times characterizing the bath are extremely short (compared to the time scales associated with the system dynamics), the presence of memory can be neglected and the Langevin Equation formalism resulting in Markovian dynamics provides an accurate description. On the other hand, if the memory times are comparable or longer than that characterizing the system of interest, the full Generalized Langevin Equation for the non-Markovian process must be analyzed. In this chapter I present a novel approximation scheme, namely the \textit{effective mass approach}, which bridges the Markovian and non-Markovian regimes and enables the analysis of systems with short memory with an approximate memoryless model. The memory effects are, however, not neglected, but are encapsulated in the effective mass of the system. This new methodology reduces the number of equations of motion describing the system and simplifies the parameter space of the model, allowing for a more comprehensive analysis of systems with short memory, especially via numerical simulations. This chapter is based on Refs.~\cite{Wisniewski2024-effmass,Wisniewski2024-entropy,Wisniewski2024-embedding}.

\section{Derivation}

Let us consider a positive, normalized and integrable memory kernel satisfying the relations
\begin{subequations} \label{eq:K:req}
\begin{align}
	\forall_{t\geq0}\, K(t) &> 0, \label{eq:K:req0}\\
	\int_0^\infty K(t) \mathrm{d} t &= 1, \label{eq:K:req1} \\ 
	\int_0^\infty t K(t) \mathrm{d} t &= \tau_K < \infty, \label{eq:K:reqtc}
\end{align}
\end{subequations}
where $\tau_K$ is expressed in time units. Since the above conditions ensure that memory must vanish at long times,~i.e., $\lim_{t\to\infty} K(t) = 0$, the quantity $\tau_K$ specifies the time scale over which the influence of past states on the system dynamics remains significant. The integrability condition in Eq.~\eqref{eq:K:reqtc} implies that in the limit $\tau_K \to 0$ the memory kernel $K(t)$ must vanish for all $t>0$. In this limit, the normalization requirement in Eq.~\eqref{eq:K:req1} is fulfilled only if $K(t)$ reduces to $K_0(t)=2\delta(t)$, representing the memoryless case. 

\begin{figure}[tb]
	\centering
	\includegraphics{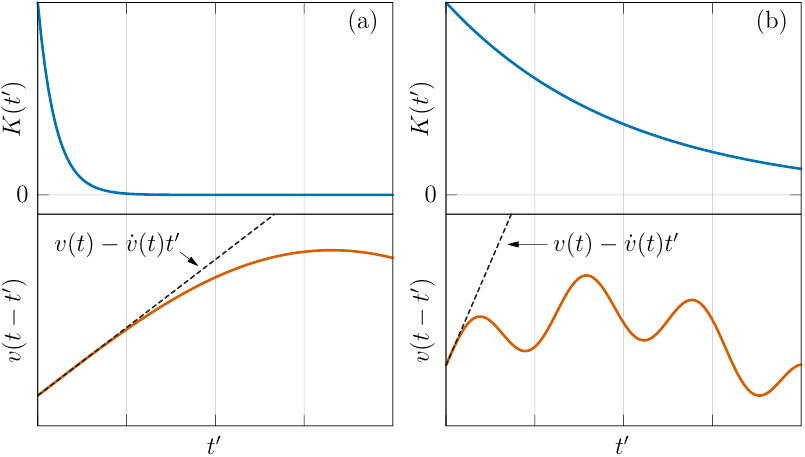}
	\caption{Schematic demonstration of the short-memory assumption. In (a) velocity $v(t-t')$ is correctly approximated with first two terms of its Taylor expansion within the time interval where the memory kernel $K(t')$ is non-negligible: the effective mass approach is expected to give correct results. In (b) the velocity strongly varies while the value of $K(t')$ is still significant: the system is out of the validity regime of the approximation.}
	\label{fig:effmass_vis}
\end{figure}

Let us now examine the friction force appearing in Eq.~\eqref{eq:GLE}
\begin{equation}
	w(t) = \gamma \int_0^t K(t-t') v(t') \mathrm{d} t'.
\end{equation}
[To be exact, in this form $w(t)$ is the friction force multiplied by $-1$].
The force $w(t)$ is a convolution of the memory kernel $K(t)$ and the velocity $v(t)$, so their arguments under the integral can be exchanged, i.e.,
\begin{equation} \label{eq:w}
	w(t) = \gamma \int_0^t K(t') v(t-t') \mathrm{d} t'.
\end{equation}
The leading contribution to the integral comes from the times $t' \lesssim \tau_K$, since for $t' \gg \tau_K$ the value of the memory kernel $K(t')$ vanishes. If $\tau_K$ is much shorter than the time scales associated with the system dynamics, one can approximate the velocity $v(t-t')$ with first two terms of its Taylor expansion around the time $t$ (see Fig.~\ref{fig:effmass_vis}),
\begin{equation}
	v(t-t') \approx v(t) - t'\dot{v}(t).
\end{equation}
Inserting the approximation into Eq.~\eqref{eq:w} gives
\begin{equation}
	w(t) \approx \gamma v(t) \int_0^t K(t') \mathrm{d} t' - \gamma \dot{v}(t) \int_0^t t'K(t') \mathrm{d} t'.
\end{equation}
For $t \gg \tau_K$ the upper limits of the integrals can be extended to infinity, so that the relations in Eqs.~\eqref{eq:K:req1} and \eqref{eq:K:reqtc} can be applied to give
\begin{equation}
	w(t) \approx \gamma v(t) - \gamma\tau_K \dot{v}(t).
\end{equation}
The first term on the right-hand side is the memoryless friction force, the same as in the Langevin Equation \eqref{eq:LE}. The other term is proportional to the acceleration $\dot{v}(t)$ and can thus be incorporated into the inertial term in Eq.~\eqref{eq:GLE}. The approximate equation for the system dynamics then reads
\begin{equation} \label{eq:EM}
	\tag{EM}
	\boxed{m^*\dot{v}(t) + \gamma v(t) = F(x, t) + \xi(t),}
\end{equation}
where
\begin{equation} \label{eq:ms}
	\boxed{m^* = m - \gamma\tau_K =  m - \Delta m}
\end{equation}
is the effective mass of the system, $\Delta m = \gamma\tau_K$ is the mass correction, and $\tau_K$ is defined in Eq.~\eqref{eq:K:reqtc}. The approximate equation \eqref{eq:EM} is identical to the memoryless Langevin Equation \eqref{eq:LE} for a system with the effective mass $m^*$. The correlated thermal noise force $\eta(t)$ was replaced with the uncorrelated one $\xi(t)$ [see Eqs.~\eqref{eq:xi0} and \eqref{eq:xitxitp}], so that the approximate system obeys the second fluctuation-dissipation relation given by Eq.~\eqref{eq:FDrel}. The form of the effective mass $m^*$ gives a hint on the range of validity of the effective mass approach. Namely, since this approximation was derived for a system with short memory, we expect the mass correction $\Delta m$ to be significantly smaller than the system mass $m$. This means that the effective mass approach is valid when $\tau_K \ll \tau_\mathrm{L}$, where $\tau_\mathrm{L} = m/\gamma$ is the velocity relaxation time of a free Brownian particle (the Langevin time). In particular, the effective mass $m^* > 0$, which ensures the dissipative character of the approximate system.

For the memory kernel $K_\mathrm{GM}(t)$ consisting of a sum of exponential decays [see Eq.~\eqref{eq:KGM}] the mass correction reads
\begin{equation}
	\Delta m_\mathrm{GM} = \gamma\sum\limits_{i=1}^n \alpha_i \tau_i.
\end{equation}
In particular, for $n=1$ the memory kernel is characterized by a single relaxation time $\tau$ and the mass correction reduces to
\begin{equation}
	\Delta m_\mathrm{M} = \gamma\tau.
\end{equation}

Summarizing, when the time $\tau_K$ characterizing the memory kernel is much shorter than the time scales associated with the system dynamics, the Generalized Langevin Equation \eqref{eq:GLE} describing non-Markovian behavior can be approximated with a memoryless Langevin Equation \eqref{eq:EM} for a system with the effective mass $m^* = m-\Delta m$. The approximate model results in Markovian dynamics, but the effects of short memory are not neglected. Instead, they are encapsulated in a correction to the system mass $\Delta m = \gamma\tau_K$, depending solely on the bath characteristics, namely the friction coefficient $\gamma$ and the time $\tau_K$.

\subsection{Second-order correction}

The class of memory kernels suitable for the application of the effective mass approach is restricted only by the requirements in Eqs.~\eqref{eq:K:req0}--\eqref{eq:K:reqtc}. Within this framework, the term $\gamma\tau_K$ can be viewed as the first-order contribution to the mass correction $\Delta m$ with respect to the time $\tau_K$. 
For the kernel $K_\mathrm{GM}(t)$ [see Eq.~\eqref{eq:KGM}]
it is possible to extend the derivation to obtain a second-order contribution. The mass correction then reads
\begin{equation} \label{eq:dm:2nd}
	\Delta m_{\mathrm{GM}} = \gamma\sum\limits_{i} \left( \alpha_i\tau_i + \frac{\alpha_i \tau_i^2}{\tau_\mathrm{L} - \sum_j \alpha_j\tau_j}\right),
\end{equation}
where $\tau_\mathrm{L}=m/\gamma$ is the Langevin time. The detailed derivation of the second-order correction can be found in Ref.~\cite{Wisniewski2024-entropy}.

\section{Validation and limitation}

\begin{figure}[tb]
	\centering
	\includegraphics{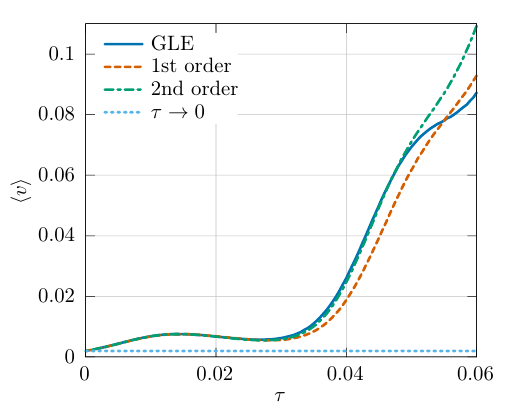}
	\caption{Average velocity $\langle v \rangle$ of the particle as a function of the memory time $\tau$. The numerical results for the Generalized Langevin Equation \eqref{eq:GLE} (solid blue) are compared with the effective mass approximation \eqref{eq:EM} with the first-order (dashed red) and second-order (dash-dotted green) mass corrections. The memoryless limit $\tau\to0$ is indicated by the light-blue dotted line. All of the considered quantities are dimensionless. Figure reproduced from Ref.~\cite{Wisniewski2024-entropy}.}
	\label{fig:effmass_demonstration}
\end{figure}

In this section I present the application of the effective mass approach to discuss its accuracy and limitations. I consider the dimensionless form of the Generalized Langevin Equation \eqref{eq:GLE:dimless} and omit the hats over the rescaled variables for clarity. For the illustrative purposes, the parameter values are set to
\begin{equation}
	m=1,\quad a=15,\quad \omega=4,\quad f=0.1,\quad \theta=10^{-3}.
\end{equation}
In the chosen dimensionless scaling the friction coefficient $\gamma$ reduces to unity (see Sec.~\ref{section:dimensionless}), and consequently the Langevin time characterizing the relaxation of a free particle is equal to the dimensionless mass, i.e., $\tau_\mathrm{L} = m = 1$. Another characteristic time is the period of the driving force $\mathsf{T}=2\pi/\omega \approx 1.57$, which is of the same order of magnitude as $\tau_\mathrm{L}$. Furthermore, I assume an exponentially decaying memory kernel $K_\mathrm{M}(t)$ characterized by a single correlation time $\tau$ [see Eq.~\eqref{eq:KM}]. Such a non-Markovian model can be approximated by the memoryless Eq.~\eqref{eq:EM} with the effective mass reading $m^* = m - \tau$. In Fig.~\ref{fig:effmass_demonstration} the average velocity of the Brownian particle $\langle v \rangle$ is presented as a function of the memory time $\tau$ for both the original system described by Eq.~\eqref{eq:GLE} and for the approximate memoryless model with an effective mass.
 
First of all, it is clear that even for memory times on the order of $\tau=0.01$, the particle velocity significantly differs from that in the memoryless limit $\tau\to0$. This indicates that the standard Langevin Equation fails to provide a suitable approximation, even though the memory time is two orders of magnitude shorter than other characteristic time scales associated with the particle's motion. The original dynamics is, however, accurately approximated if the mass correction defined in the effective mass approach is taken into account. 

Specifically, the approximate model correctly reproduces the particle dynamics for memory times up to $\tau \approx 0.03$ when the first-order correction is applied, and up to $\tau \approx 0.05$ if the second-order correction is included. As the memory time increases, the condition $\tau \ll \tau_\mathrm{L}$ is no longer satisfied, and the system moves beyond the regime of validity for the effective mass approach.

\section{Comparison with Markovian embedding}

Both the effective mass approach and the Markovian embedding procedure allow for the description of non-Markovian dynamics within a Markovian framework. In this section, I contrast these two methodologies, highlighting their advantages and discussing how they may be combined.

The primary advantage of the Markovian embedding method is its accuracy, namely, the set of equations \eqref{eq:embedding:set} is mathematically equivalent to the Generalized Langevin Equation \eqref{eq:GLE}. In other words, Markovian embedding is not an approximation, but rather a different formulation of the original problem. However, its applicability is limited to memory kernels representable in the form of Eq.~\eqref{eq:embedding:K}. Here lies the strength of the effective mass approach, which accommodates a significantly broader class of memory kernels, including Gaussian $K_\mathrm{G}(t) \propto \exp\left[-(t/\tau)^2\right]$ and algebraic $K_\mathrm{A}(t) \propto (t+\tau)^{-k}$ for $k>2$.

Another key distinction is the computational complexity. The Markovian embedding method requires the solution of $n$ additional equations of motion for $n$ auxiliary variables, and generation of $n$ corresponding noise sources. As a result, in the latter method the computational time scales as $\mathcal{O}(n)$, which can limit the capabilities of exploring the parameter space of the model. This problem does not occur in the effective mass approach, whose complexity is comparable to that of the memoryless Langevin Equation formalism and is independent on the particular form of $K(t)$, since the mass correction $\Delta m$ can be typically calculated analytically before the simulations.

Finally, a memory kernel composed of a sum of exponential decays can serve as a flexible approximation for other memory kernels, such as $K_\mathrm{A}(t)$. Given $n$ exponential terms and the normalization condition in Eq.~\eqref{eq:K:req1}, the model contains $2n-1$ adjustable parameters $\tau_i$ and $\alpha_i$, offering a high degree of freedom. However, for short memory times, the accuracy of the effective mass approach might be comparable to that of this more complex scheme, despite its simplicity. For longer memory times, the more sophisticated approximation basing on Markovian embedding offers superior accuracy, provided that the parameters are appropriately tuned. Notably, optimal accuracy is achieved when the parameters are chosen such that both the original and approximate kernels yield identical mass corrections within the effective mass framework. This tuning strategy has proven superior to traditional least-squares fitting of the kernels, which I showed in Ref.~\cite{Wisniewski2024-embedding}.

\chapter{Dynamics of non-Markovian systems with short memory \label{chapter:transport}}
\chaptermark{Dynamics of non-Markovian systems}

In this chapter I present three examples illustrating how short memory influences the dynamics of non-Markovian systems. The first two cases focus on the direction of the particle's motion, which is of critical significance for applications such as target-seeking or particle sorting. The last example demonstrates how short memory can impact the stability of the running solutions and proposes a potential application for the observed effect. Collectively, these examples indicate the usefulness of the effective mass approach, which was used to discover these phenomena and provided a physical interpretation of their origins. In the following sections, the Generalized Langevin Equation is analyzed in the dimensionless form [see Eq.~\eqref{eq:GLE:dimless}], with the hats over the rescaled variables omitted for simplicity.

\section{Memory-induced absolute negative mobility \label{section:memory-inducedANM}}

When a dynamical system is subjected to a weak external perturbation, its response is expected to be proportional to the applied forcing, indicating that the system remains within the linear response regime. If the external perturbation is the constant force $f$ and the response is quantified with the average velocity $\langle v \rangle$, the proportionality constant is the absolute mobility $\mu_0$, as defined in Eq.~\eqref{eq:mu0}. 

In linear systems, the superposition principle dictates that the contributions of individual forces to the total response can be analyzed independently. In particular, forces with a vanishing mean [such as the periodic driving $a\cos(\omega t)$ or the thermal noise $\eta(t)$] do not contribute to the average velocity $\langle v \rangle$. Consequently, the particle follows the direction of the external bias $f$, and the absolute mobility must be positive. Furthermore, if the system is near the thermodynamic equilibrium, the Le Chatelier-Braun principle states that the system response to external perturbation counteracts the disturbance \cite{Machura2007,Speer2007}, which again implies $\mu_0 > 0$.

However, in the system considered in this chapter, with the external force $F(x, t)$ given by Eq.~\eqref{eq:Fxt}, neither of these constraints apply. First, the presence of a periodic potential [either $V_\mathrm{S}(x)$ in Eq.~\eqref{eq:Vs} or $V_\mathrm{R}(x)$ in Eq.~\eqref{eq:Vr}] introduces non-linearity, and consequently the superposition principle cannot be applied. Second, the periodic driving force $a\cos(\omega t)$ ensures that the system is maintained far from equilibrium, which invalidates the Le Chatelier-Braun principle. Under these conditions, the particle does not necessarily follow the direction of the external perturbation, allowing for the emergence of \textit{absolute negative mobility} (ANM), where $\mu_0 < 0$.

The existence of ANM has been confirmed both theoretically (see e.g.~Ref.~\cite{Wisniewski2022-ANM} and references therein) and experimentally \cite{Ros2005,Nagel2008,Luo2016}. This counterintuitive phenomenon arises in both the stochastic dynamics of Brownian particles and the deterministic (zero-temperature) limit $T\to 0$ \cite{Machura2007, Wisniewski2023-weakdiss}. In this section, I present a case where ANM occurs exclusively when (short) memory of the thermal bath is taken into account and vanishes in the memoryless limit \cite{Wisniewski2024-memoryANM}.

\begin{figure}[tb]
	\centering
\begin{minipage}[t]{0.475\textwidth}
	\centering
	\includegraphics{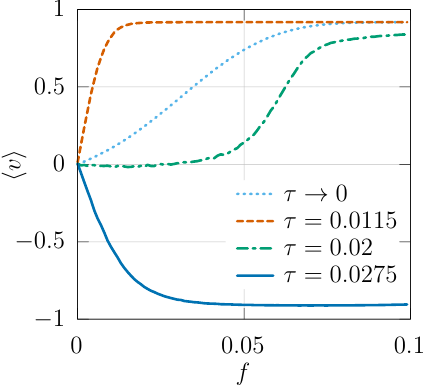}
	\caption{Average velocity of the particle $\langle v \rangle$ as a function of the static bias $f$ for different memory times $\tau$. Figure reproduced from Ref.~\cite{Wisniewski2024-memoryANM}.}
	\label{fig:v_f}
\end{minipage}
\hfill
\begin{minipage}[t]{0.475\textwidth}
	\centering
	\includegraphics{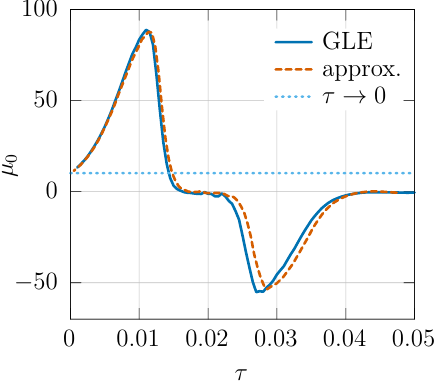}
	\caption{Absolute mobility $\mu_0$ as a function of the memory time $\tau$. Figure reproduced from Ref.~\cite{Wisniewski2024-memoryANM}.}
	\label{fig:mu_tm}
\end{minipage}
\end{figure}

Let us consider the dynamics of a Brownian particle in a non-Markovian bath with an exponentially decaying memory kernel $K_\mathrm{M}(t)$, placed in a symmetric sinusoidal potential $V_\mathrm{S}(x)$. Throughout this section, I adopt the following parameter regime:
\begin{equation}
	m = 0.907,\quad a=12.2,\quad \omega = 5.775,\quad \theta = 10^{-4}.
\end{equation}
Fig.~\ref{fig:v_f} illustrates the average velocity of the particle $\langle v \rangle$ as a function of the static bias $f$ for various memory times $\tau$.
In the linear response regime, the mobility $\mu_0$ corresponds to the slope of the average velocity curve $\langle v \rangle(f)$ in the limit of vanishing bias ($f \to 0$). In the Markovian limit ($\tau \to 0$), the mobility is positive ($\mu_0 > 0$). As the correlation time $\tau$ increases, the slope initially steepens. However, at $\tau = 0.0275$ the mobility becomes strictly negative ($\mu_0 < 0$), indicating the onset of absolute negative mobility. This demonstrates that, within the evaluated parameter regime, ANM is manifested solely due to the presence of non-zero memory. Note that the characteristic times associated with the particle dynamics, such as the Langevin time $\tau_\mathrm{L} = 0.907$ and the period of the driving $\mathsf{T} \approx 1.09$, are two orders of magnitude longer than the memory time $\tau$ for which ANM occurs. The absolute negative mobility phenomenon induced exclusively by very short memory has not been previously reported.

To understand the origin of this memory-induced ANM, let us apply the effective mass approach to the considered non-Markovian model. For the exponentially decaying memory kernel, the mass correction with the second order term reads
\begin{equation} \label{eq:KM:dm}
	\Delta m = \tau + \frac{\tau^2}{m-\tau}
\end{equation}
[see Eq.~\eqref{eq:dm:2nd} and recall that in the considered scaling the friction coefficient $\gamma$ reduces to unity]. The emergence of the negative mobility in the original non-Markovian system should thus be accompanied by a similar effect in the Markovian model with an effective mass. This is confirmed in Fig.~\ref{fig:mu_tm}, which compares the $\mu_0(\tau)$ curves obtained for the original Eq.~\eqref{eq:GLE} and the approximate model described by Eq.~\eqref{eq:EM}. The effective mass approach yields correct results for $\tau \lesssim 0.01$. For longer memory times, the quantitative agreement diminishes, but the qualitative behavior remains consistent. Crucially, in both models there is a range of memory times $\tau$ (corresponding also to mass corrections $\Delta m$) for which $\mu_0(\tau)<0$. Consequently, the emergence of ANM in the non-Markovian system can be explained by the presence of a similar phenomenon in the approximate Markovian model with an effective mass.

The results show that even very short memory can radically change the behavior of a physical system and lead to counterintuitive phenomena, such as the absolute negative mobility. This fact must be contrasted with the common assumption that the presence of short memory can be safely neglected and no memory effects are expected to be observed. In this section I demonstrated that the Markovian approximation must be applied with special caution and some corrections, such as those appearing in the effective mass approach, may be necessary to take into account the effects of short memory.

\section{Memory-induced current reversal}

Average directed transport is fundamentally prohibited in spatially symmetric systems, in which every trajectory is accompanied by its counterpart with an opposite velocity, causing their contributions to the net transport to cancel out \cite{Denisov2014}. While in Sec.~\ref{section:memory-inducedANM} the spatial symmetry was broken by the constant bias $f$, in the absence of this external force asymmetry may instead be introduced through the potential $V(x)$.

A classic example of such an asymmetric landscape is the ratchet potential $V_\mathrm{R}(x)$, characterized by alternating steep and gentle slopes [see Eq.~\eqref{eq:Vr} and Fig.~\ref{fig:potentials}(b)]. The direction of transport in such ratchet systems is typically difficult to predict \textit{a priori} and current reversal, i.e., a change in the direction of motion, can be observed upon variation of parameters characterizing, e.g., the potential profile \cite{Chauwin1995,Jiao2023}, external forcing \cite{Jung1996,Mateos2000} and thermal noise intensity \cite{Kula1998a}.

Ratchet potentials are frequently employed to model intracellular transport, where biological motors navigate on asymmetric substrates (microtubules) \cite{Hanggi2009}. Given that the intracellular environment is viscoelastic, the dynamics of these motors is expected to exhibit memory effects \cite{Goychuk2010}. In this section, I demonstrate that memory serves as a critical factor influencing the direction of the motor's motion. Specifically, I show that in a correlated environment, the net transport of Brownian particles can be opposite to that in its memoryless counterpart \cite{Wisniewski2025-currrev}.

To model the asymmetric landscape, I utilize the potential $V_\mathrm{R}(x)$ [see Eq.\eqref{eq:Vr}]. Since the spatial symmetry in this potential is already broken, the presence of a constant bias is no longer necessary to observe directed transport, therefore I consider the case $f=0$. In this section, I adopt the following parameter regime:
\begin{equation}
	m = 0.525,\quad a = 17,\quad \omega = 11,\quad D = 10^{-3},
\end{equation}
for which in the memoryless limit $\tau \to 0$ the particle's average velocity is positive, i.e., $\langle v \rangle > 0$.
\begin{figure}[tb]
	\centering
	\includegraphics{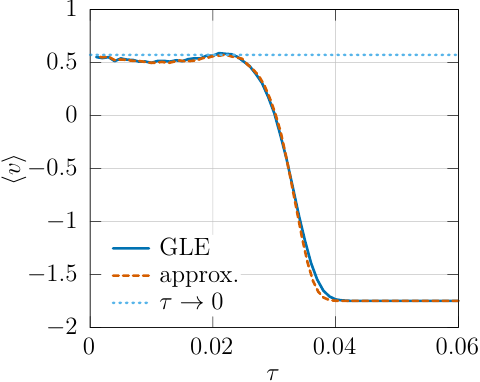}
	\caption{Average velocity of the particle $\langle v \rangle$ as a function of the memory time $\tau$ for the non-Markovian system modeled with Generalized Langevin Equation \eqref{eq:GLE} and its Markovian counterpart with effective mass described with Eq.~\eqref{eq:EM}. Figure reproduced from Ref.~\cite{Wisniewski2025-currrev}.}
	\label{fig:v_tm}
\end{figure}
Figure~\ref{fig:v_tm} illustrates the influence of memory on particle transport by plotting $\langle v \rangle$ against the memory time $\tau$. In the memoryless limit, the current is positive and remains so for $\tau \lesssim 0.03$. However, as $\tau$ increases, $\langle v \rangle$ changes sign, clearly demonstrating the current-reversal effect.

The physical origin of this phenomenon can be elucidated through the effective mass approximation. As shown in Fig.~\ref{fig:v_tm}, current reversal also emerges in the Markovian counterpart of the studied system when the bare mass $m$ is replaced by an effective mass $m^* = m - \Delta m$, where $\Delta m$ is again given by Eq.~\eqref{eq:KM:dm}. Specifically, the figure reveals an overlap between the dynamics of the original model and that in the effective mass approach. This indicates that memory \textit{effectively} reduces the mass of the system, pushing it into a parameter regime where the current-reversal phenomenon naturally arises.
\begin{figure}[tb]
	\centering
	\begin{minipage}[t]{0.48\textwidth}
		\centering
		\includegraphics{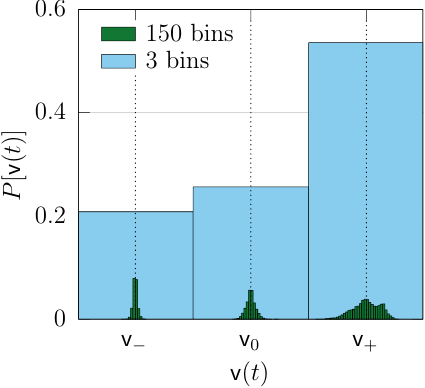}
		\caption{Probability distribution $P[\mathsf{v}(t)]$ in the memoryless limit $\tau\to0$ calculated from histograms consisting of 150 and 3 bins. The values of the three bins correspond to the probabilities $P_\pm$ and $P_0$. Figure reproduced from Ref.~\cite{Wisniewski2025-currrev}.}
		\label{fig:histogram}
	\end{minipage} \hfill
	\begin{minipage}[t]{0.48\textwidth}
		\centering
		\includegraphics{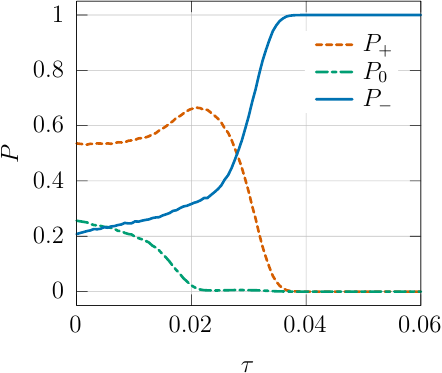}
		\caption{The probabilities $P_\pm$ and $P_0$ for the emergence of the states $\mathsf{v}_{\pm} = \pm\omega/(2\pi)$ and $\mathsf{v}_0 = 0$, respectively, as a function of the memory time $\tau$. Figure reproduced from Ref.~\cite{Wisniewski2025-currrev}.}
		\label{fig:probability}
	\end{minipage}
\end{figure}

The memory-induced current reversal effect can be further understood by analyzing the particle trajectories in the phase space of the system. For each trajectory, one can calculate the period-averaged velocity $\mathsf{v}(t)$ according to Eq.~\eqref{eq:vperiod}. The net average velocity of the particle can then be expressed as the long-time limit of the ensemble-averaged $\mathsf{v}(t)$ \cite{Jung1993}, i.e.,
\begin{equation}
	\langle v \rangle = \lim_{t \to \infty} \langle \mathsf{v}(t) \rangle.
\end{equation}
Fig.~\ref{fig:histogram}, presents the probability distribution $P[\mathsf{v}(t)]$ in the long-time limit for the memoryless model.
The velocity $\mathsf{v}(t)$ is distributed around three distinct values corresponding to the attractors present in the deterministic system with $\theta = 0$: two running solutions $\mathsf{v}_\pm = \pm\omega/(2\pi)$, in which the particle covers one spatial period per driving cycle $\mathsf{T}$, and a locked state $\mathsf{v}_0 = 0$, where the particle's motion is confined to a single potential well.

In the presence of thermal noise, the particle randomly switches between these states, however, in the long-time limit, the probability distribution $P[\mathsf{v}(t)]$ converges to a time-invariant measure. Fig.~\ref{fig:probability} displays the occupation probabilities $P_\pm$ and $P_0$ corresponding to the states $\mathsf{v}_\pm$ and $\mathsf{v}_0$ as a function of the memory time $\tau$. In the memoryless limit, the majority of trajectories occupy the positive velocity state $\mathsf{v}_+$. As $\tau$ increases, the negative velocity state $\mathsf{v}_-$ quickly becomes more populated, eventually saturating at $P_- = 1$. At this point, the probability of escaping the $\mathsf{v}_-$ state becomes negligible. This reveals the mechanism behind the emergence of the current-reversal phenomenon: it is rooted in a memory-induced dynamical localization effect \cite{Spiechowicz2017,Spiechowicz2019}, wherein all system trajectories are funneled into the negative velocity attractor as the correlation time grows.

The presented phenomenon reveals another aspect of the possible influence of short memory on the system dynamics. Namely, even though in the short-memory regime the character of the underlying attractors remains identical to that in the memoryless case (at least with respect to the period-averaged velocity), their occupation changes, which affects the direction of the net transport. Remarkably, a similar scenario is observed in a corresponding memoryless setup upon variation of the system mass, confirming the validity of the effective mass approach.

\section{Memory-controlled random bit generator}

A standard assumption in the analysis of non-Markovian systems is that the memory time characterizing the surroundings remains constant. There are, however, setups whose properties can be modified with external stimuli, such as light \cite{Marozas2019,Carberry2020,Lu2025} or an electric field \cite{Teng2024}. This capability allows the characteristic quantities, such as fluidity and memory time, to be tuned ``on the fly,'' converting them into active control parameters. The ability to modify the system properties on demand opens up new possibilities for steering the dynamics of immersed microscopic objects. In this section, I propose a setup that leverages the tunable viscoelasticity to control the generation and storage of binary information \cite{Wisniewski2025-randombit}.

Specifically, I present a system in which the memory time characterizing the thermal bath dictates whether the particle resides in one of two stable states representing bits of information or enters a chaotic regime where previous information is erased and a new bit value is generated. Traditionally, such information is encoded in the spatial position of a Brownian particle within a double-well potential, where each well represents a distinct logical state \cite{Parrondo2015}. In contrast, the methodology presented here encodes information in the period-averaged velocity $\mathsf{v}(t)$, which can assume either positive or negative values. I adopt the convention that these states correspond to logical ``1'' and ``0'' bits, respectively.

To ensure that both bits are equally probable, I consider a symmetric model featuring a sinusoidal potential $V_\mathrm{S}(x)$ and no external bias $f=0$. While the macroscopic net transport in such a system strictly vanishes, the binary information is instead carried by the velocity of a single particle. For an individual particle, the spatial symmetry can be broken by the choice of initial conditions, namely its position $x(0)$, velocity $v(0)$, and the initial phase of the driving force $a\cos(\omega t)$. Unless stated otherwise, the following parameter set is utilized throughout this section:
\begin{equation}
	m = 1,\quad a = 8,\quad \omega = 5,\quad \theta = 10^{-4},
\end{equation}
though the underlying operational principle remains general.

\begin{figure}[bt]
	\centering
	\includegraphics{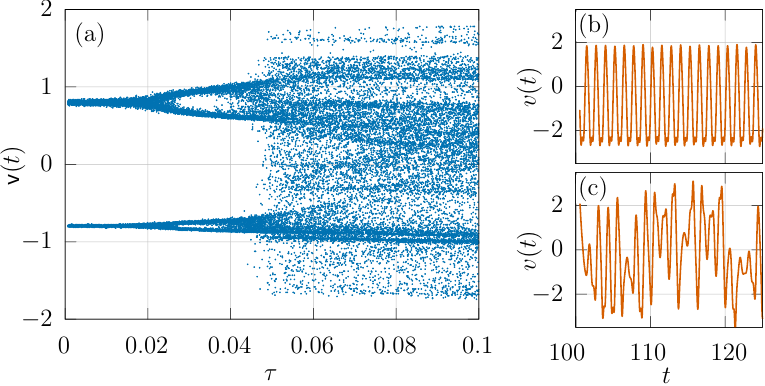}
	\caption{(a) ``Bifurcation'' diagram of the period-averaged velocity $\mathsf{v}(t)$ as a function of the memory time $\tau$. Panels (b) and (c) depict exemplary trajectories for the bistable ($\tau=0.01$) and chaotic ($\tau = 0.1$) regimes, respectively. Figure reproduced from Ref.~\cite{Wisniewski2025-randombit}.}
	\label{fig:bifurc_tau}
\end{figure}

The analysis begins with a ``bifurcation'' diagram of the period-averaged velocity $\mathsf{v}(t)$ as a function of the memory time $\tau$, presented in Fig.~\ref{fig:bifurc_tau}(a). 
This diagram was generated by solving the equations of motion for various initial conditions and realizations of the thermal noise $\eta(t)$ over $10^3$ periods of the driving force $\mathsf{T}$. The period-averaged velocity $\mathsf{v}(t)$ was subsequently calculated for each trajectory by averaging the instantaneous velocity $v(t)$ over the final driving period.

For short memory times $\tau \lesssim 0.02$ (including the memoryless limit $\tau = 0$), the period-averaged velocity converges to two discrete values, $\mathsf{v}_{\pm} = \pm \omega/(2\pi) \approx \pm 0.8$, which are slightly broadened by the presence of thermal noise. This bimodal distribution arises because $v(t)$ is strictly periodic with period $\mathsf{T}$ [see Fig.~\ref{fig:bifurc_tau}(b)]. The values $\mathsf{v}_\pm$ correspond to stable running solutions where the particle advances by exactly one spatial period of the potential $V_\mathrm{S}(x)$ during each driving cycle, moving in either the positive or negative direction. The absence of intermediate points indicates that noise-induced switching between these attractors is exceedingly rare.

In stark contrast, for $\tau \gtrsim 0.05$, the period-averaged velocity spans nearly the entire continuous range between $-1.8$ and $1.8$. In this regime, the instantaneous velocity $v(t)$ becomes aperiodic [see Fig.~\ref{fig:bifurc_tau}(c)], causing its period average to fluctuate significantly depending on the initial conditions and the specific observation window. Consequently, this system perfectly fulfills the operational requirements for a random bit generator. Namely, in the short-memory regime, the particle reliably settles into one of two stable states representing values of an information bit. For longer memory times, the velocity dynamics loses the regularity, allowing this chaotic state to be exploited as a generator for random bit values. 

\begin{figure}[tb]
	\centering
	\includegraphics{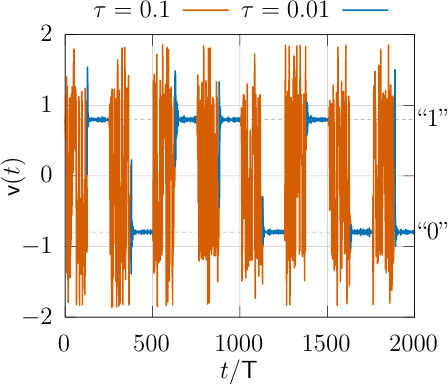}
	\caption{Time evolution of the period-averaged velocity $\mathsf{v}(t)$ under periodic switching of the memory time $\tau$ between $0.01$ and $0.1$ every $100$ periods of the driving force $\mathsf{T}$. The corresponding bit sequence is ``$10110100$''. Figure reproduced from Ref.~\cite{Wisniewski2025-randombit}.}
	\label{fig:traj_switch}
\end{figure}

The operational protocol of the proposed random bit generator is illustrated in Fig.~\ref{fig:traj_switch}. The memory time $\tau$ is periodically switched every $100$ periods of the driving force $\mathsf{T}$ between $\tau = 0.1$ (the ``chaotic state'') and $\tau = 0.01$ (the ``bistable state''). Consequently, the period-averaged velocity $\mathsf{v}(t)$ alternates between irregular behavior and one of the two stable states $\mathsf{v}_\pm$. As demonstrated in Ref.~\cite{Wisniewski2025-randombit}, the successive bits generated by this process are statistically random provided the generation time exceeds $5\mathsf{T}$. Furthermore, I have confirmed that the bistable behavior for $\tau \lesssim 0.02$ remains robust as long as the noise intensity satisfies $\theta \ll 0.02$, a threshold corresponding to the effective energy barrier separating the bistable attractors.

The presented methodology provides a general operational principle for designing similar systems for storing and processing information on a microscopic scale. Besides the period-averaged velocity, the information may also be stored in other dynamical quantities, such as the angular velocity for rotary objects. The method can be applied whenever the dynamics changes from regular to chaotic upon variation of the memory time. In the short-memory case, the suitable parameter regime can also be found by analyzing the approximate memoryless model with the effective mass.

\chapter{Concluding remarks \label{chapter:conclusions}}

The dynamics of physical systems is inherently non-Markovian, which frequently stems from the reduction of the number of degrees of freedom describing either the system or its surroundings. This fact is especially pronounced at the mesoscopic level, where the interaction of the system with its environment lies at the core of the observed phenomenology. Although it is frequently assumed that the Markovian approximation neglecting any temporal correlations is justified when the memory time is significantly shorter than the characteristic times associated with the system dynamics, it remains unclear what degree of time scale separation is required for its validity.
In this dissertation, I have focused on characterizing the short-memory regime and demonstrated that even short temporal correlations may have pronounced impact on the system dynamics.

First, I showed that the behavior of systems with short memory is approximately equivalent to that of memoryless setups with an adjusted mass.
The presented effective mass approach simplifies the parameter space of the model and reduces the number of equations of motion required for its description, which effectively enables a more thorough analysis, especially with the use of computational methods. Furthermore, this novel methodology offers an intuitive interpretation of the origin of various memory-induced effects, as similar phenomena occur in memoryless setups upon variation of the system mass.

Building upon this new theoretical framework, I demonstrated that memory can play a pivotal role in transport at microscopic scale, as its presence may reverse the direction of motion of Brownian particles in periodic potentials. Notably, such reversal may occur even when the memory time is two orders of magnitude shorter than the characteristic times associated with the system dynamics. These findings show that even when these time scales appear clearly separated, the memory effects may nevertheless remain remarkably pronounced. This means that the frequently applied Markovian approximation should be applied with special caution. Future research could investigate how memory affects the emergence of directed transport, i.e., whether its presence may be necessary for the directed motion to occur. Moreover, the impact of anticorrelations arising when the memory kernel takes negative values remains to be determined.

Furthermore, I showed that the sensitivity of the system dynamics to the characteristics of its thermal bath provides an opportunity to control the system by tuning the memory time. In particular, I proposed a setup, in which the memory time acts as a control parameter for the generation and storage of information bits. In the presented methodology, information is stored in the period-averaged velocity of a Brownian particle, however, this operational principle can be generalized to other observables, given the broad class of universality of Brownian motion. The presented results demonstrate that the ability to manipulate the memory time, already attainable experimentally, opens doors to novel applications and provides new possibilities for the engineering of microscopic systems. Further studies could elaborate similar methodology in the overdamped (inertialess) limit, which is frequently the focus of experimental research.

Collectively, the findings presented in this dissertation demonstrate the importance of short-time correlations in the dynamics of mesoscopic systems. All the presented findings can be verified experimentally e.g.~in a setup consisting of colloids immersed in a viscoelastic medium and manipulated with optical tweezers, but the results can be generalized for a broad class of problems, given that the dynamics of Brownian particles is of fundamental meaning in physics, chemistry, biology and engineering. Ultimately, since ``non-Markov is the rule, Markov is the exception,'' as Nico van Kampen famously noted \cite{Van_Kampen1998}, memory occurs universally in all these fields. The recent advancements in the field of non-Markovian dynamics (see \hyperlink{target:discoveries}{references in the Introduction}) suggest that there is still a wealth of phenomena awaiting discovery, and accounting for the memory effects will deepen our understanding of nature.

\backmatter

\phantomsection
\addcontentsline{toc}{chapter}{Bibliography}

\newrefcontext[labelprefix={}]
\printbibliography[notkeyword={phd},notkeyword={additional}]

\end{document}